\documentclass[aps,prb,twocolumn,epsfig,reprint,amsmath,groupedaddress,amssymb,float]{revtex4-1}

\usepackage{graphics}
\usepackage{graphicx}
\usepackage{epsfig}
\usepackage{subfigure}
\usepackage{amssymb}
\usepackage[dvips]{color}

\begin{document}

\preprint{}

\title{Electron-phonon coupling, superconductivity and nontrivial band topology in NbN polytypes}

\author{K. Ramesh Babu and Guang-Yu Guo}
\affiliation{Department of Physics and Center for Theoretical Physics, National Taiwan University, Taipei 10617, Taiwan\\
Physics Division, National Center for Theoretical Sciences, Hsinchu 30013, Taiwan\\
}

\date{\today}

\begin{abstract}
In this paper, we investigate the mechanical properties, electronic band structure, lattice dynamics
and electron-phonon interaction in $\delta$-NbN, $\varepsilon$-NbN, WC-NbN and $\delta^\prime$-NbN by performing systematic {\it ab initio}
calculations based on density functional theory with the generalized gradient approximation.
We find that all the four structures are mechanically stable
with $\varepsilon$-NbN being the ground state structure. The calculated elastic constants, which agree well with
available experimental data, demonstrate that all four NbN polytypes are hard materials with
bulk modulii being close to that of boron nitride. The calculated electronic
band structures show that all four polytypes are metallic with the Nb \textit{d}-orbital
dominated energy bands near the Fermi level ($E_F$).
The calculated phonon dispersion relations of $\delta$-NbN are in good agreement with neutron scattering experiments.
The electron-phonon coupling ($\lambda$)
in $\delta$-NbN ($\lambda=0.98$) is much stronger than in $\varepsilon$-NbN ($\lambda=0.16$), WC-NbN ($\lambda=0.11$) and $\delta^\prime$-NbN ($\lambda=0.17$). This results in a much higher superconducting
transition temperature ($T_c =18.2$ K) than in $\varepsilon$-NbN, WC-NbN and $\delta^\prime$-NbN ($T_c \le 1.0$ K).
The stronger $\lambda$ and higher $T_c$ in $\delta$-NbN are attributed to its large density of states at $E_F$
and small Debye temperature. The calculated $T_c$ of $\delta$-NbN is in good agreement with the experimental values.
However, the predicted $T_c$ of $\varepsilon$-NbN is much smaller than the recent experiment (11.6 K) but
agrees well with the earlier experiment, suggesting further experiments on single phase samples.
Finally, the calculated relativistic band structures reveal that all four NbN polytypes are
topological metals. Specifically, $\varepsilon$-NbN and $\delta^\prime$-NbN are type-I Dirac metals whereas $\delta$-NbN is type-II
Dirac metal, while WC-NbN is an emergent topological metal that has rare triply degenerate nodes.
All these results indicate that all the four NbN polytypes should be hard superconductors with nontrivial band topology
that would provide valuable opportunities for studying fascinating phenomena arising from
the interplay of band topology and superconductivity.
\end{abstract}


\maketitle

\section{INTRODUCTION}
Ever since superconductivity was discovered in early 20th century by Kammerlingh Onnes, the quest for discovering materials
with higher transition temperature has always been challenging and hence has continued receiving unrelenting attention.
Excitingly, hydrogen sulfide was recently demonstrated to have a superconducting transition
at 203 K under high pressure.\cite{Drozdov15} Furthermore, yttrium-based hydrogen clathrate structures were predicted to
exhibit room temperature superconductivity with a transition temperature of 303 K under  pressure of 400 GPa.\cite{Feng}
In the meantime, materials that show both topological properties and superconductivity have recently received intensive
interest because of possible realization of exotic Majorana fermions, particles coinciding with their own anti-particles,
in such condensed matter systems.\cite{Angus} Therefore, it is highly motivated to investigate the materials
that have topological properties and superconductivity. In this paper, we pay attention to transition metal nitrides (TMNs)
in which some of them are good superconducting materials with transition temperature ranging from $T_c=8$ K
for TaN \cite{Warren} to $T_c=17$ K for $\delta$-NbN \cite{Ralls, LEToth} and others are topological metals \cite{Bian}.

It is well known that TMNs are good candidates for technological applications because
of their superior mechanical, electronic and superconducting properties.\cite{Pierson} They have been widely used
as microelectronic devices, protective-resistant coatings, high pressure devices etc.\cite{XJChen} While most of the TMNs
crystallize in a cubic structure, there are, for example, group V nitrides with different polymorphic
structures.\cite{GBrauer1,GBrauer,NTerao,Gold,Terao} Out of group V TMNs, niobium nitride NbN has four polymorphic
structures, namely, cubic NbN ($\delta$-NbN, NaCl structure) \cite{GBrauer1,GBrauer},
hexagonal NbN [AsNi type ($\delta^\prime$-phase) \cite{NTerao}; tungsten carbide (WC) type \cite{Gold}
and $\varepsilon$-phase \cite{Terao}]. Due to its notable mechanical properties and existence in various polytypes,
NbN has received considerable attention in recent years.\cite{YZou,YZou1,Anand} Chen {\it et al.} studied the mechanical properties of $\delta$-NbN by Vickers indentation method and found that it has bulk modulus comparable to that of hard materials such as
cubic boron nitride and close to that of sapphire.\cite{XJChen} The electronic structure was studied theoretically
by means of plane-wave non-local pseudopotential method\cite{DJChadi}, linear muffin-tin orbital method \cite{BPalanivel}
and linearized augmented plane wave method.\cite{KSchwarz,TAmriou} Christensen {\it et al.} measured the phonon dispersion
of $\delta$-NbN$_{0.93}$ by in-elastic neutron scattering and found that anomalies exist in acoustic phonon branches
at X-point in the Brillouin zone. \cite{Christ} Theoretical studies based on density functional theory revealed
that the phonon dispersion of $\delta$-NbN indeed shows soft modes at X point which lead to lattice instability of
the structure similar to other nitrides such as VN and HfN in their cubic form.\cite{EIIsaev,EIIsaev1,Olifan,Blackburn,Jha}
However, {\it ab initio} calculations on the superconducting properties of $\delta$-NbN are relatively less addressed
in the literature compared to other transition metal nitrides.\cite{BPalanivel,EIIsaev,YZou}
Here we perform systematic ab inito calculations on the mechanical properties, electronic structure, lattice dynamics,
electron-phonon coupling and superconducting properties of $\delta$-NbN.

The crystal structure of hexagonal $\varepsilon$-NbN was first reported by Terao {\it et al.}~\cite{Terao} In light of
superconductivity of $\delta$-NbN, it is prevalent to search for superconductivity in $\varepsilon$-NbN. However,
the experiments by Oya {\it et al.} \cite{Oya} concluded that the hexagonal $\varepsilon$-NbN does not exhibit superconductivity
above 1.77 K. On the other hand, the recent experiments on magnetization and electrical resistivity of $\varepsilon$-NbN
claimed the existence of superconductivity with transition temperature as high as $\sim$ 11.6 K.\cite{YZou1}
Clearly, there is a controversy about the superconductivity in $\varepsilon$-NbN, which thus requires theoretical calculations
for better understanding of the system as well as further experiments. In the present study, we aim to study
the superconducting properties of $\varepsilon$-NbN by performing {\it ab initio} density functional theory calculations.
The electronic band structure, density of states and phonon dispersion of $\varepsilon$-NbN are calculated which are then used
to study the electron-phonon coupling and superconductivity in $\varepsilon$-NbN. Moreover, the ultrasonic experiments
on $\varepsilon$-NbN revealed that the material has superior mechanical properties compared to $\delta$-NbN and
the structure is stable up to pressures of 20 GPa.\cite{YZou1} Therefore, we also calculate the elastic properties
of $\varepsilon$-NbN and compare them with that of $\delta$-NbN.

Very recently, Chang {\it et al.} \cite{Bian} predicted that materials that crystallize in the hexagonal WC structure,
could host an exotic topological phase that goes beyond Dirac and Weyl semimetals and features triply-degenerate nodal points
along the $k_z$ direction ($\Gamma$-A symmetry lines) in the Brillouin zone. Indeed, these triply-degenerate nodal points
were recently observed experimentally in topological semimetal MoP.\cite{Lv17}
Since NbN also exists in WC structure and WC structure was found to be energetically more stable than the cubic
structure $\delta$-NbN\cite{Olifan, ZH}, WC-NbN could well be another material that could show the new topological properties. \cite{Bian,Ziming}
Therefore, it would be worthwile to carry out a detailed analysis of the electronic band structure of WC-NbN.

There are some theoretical reports available on the phonon dispersion\cite{EIIsaev1,Olifan} and electronic structure
of WC-NbN \cite{ZH,Litinskii}. However, \textit{ab initio} studies on the superconducting properties of the hexagonal WC-NbN
are still lacking in the literature.\cite{EIIsaev1,Olifan,Litinskii,ZH} Furthermore, NbN also exists in 
hexagonal anti-NiAs type structure ($\delta^\prime$-NbN)\cite{NTerao} which received less attention in the previous literature.\cite{YZou,YZou1} In particular, there are no theoretical studies available on the electronic structure, elastic and superconducting properties of this structure.\cite{CWang,EIIsaev1,Olifan} Therefore, a comparative study of the superconducting
properties of all the four polytypes of NbN is required. In the present work, we systematically investigate
the electronic structure, mechanical properties, phonon dispersion and electron-phonon interactions of all the four polytypes
of NbN by performing \textit{ab initio} calculations. In general, for conventional superconductors
with dominant electron-phonon interactions, the superconductivity properties can be analyzed through calculating
the Eliashberg spectral function $\alpha^2F(\omega)$. Hence we calculate the $\alpha^2F(\omega)$ function along
with phonon dispersion and phonon density of states for the four polytypes. Then, by using Allen-Dynes formula
the superconducting transition temperatures for the four structures are calculated.

The rest of this paper is organized as follows. In section II, we introduce the crystal structures and also theory
of superconductivity along with the theoretical methods and computational details used in the present study.
The calculated physical properties of these NbN polytypes are presented and analyzed in section III.
Finally, we give a summary of the conclusions drawn from the present work in section IV.

\section{CRYSTAL STRUCTURES AND COMPUTATIONAL METHODS}
The crystal structure of $\delta$-NbN is cubic with Fm$\bar{3}$m space group
and contains one formula unit (f.u.) per cell.~\cite{GBrauer1,GBrauer}
Nb occupies the position (0, 0, 0) and N is at (1/2, 1/2, 1/2). $\varepsilon$-NbN~\cite{Terao}, WC-NbN~\cite{Gold} and $\delta^\prime$-NbN~\cite{NTerao}
crystallize in hexagonal structures with P6$_3$/mmc, P$\bar{6}$m2 and P6$_3$/mmc space group, respectively.
The unit cell of $\varepsilon$-NbN contain
four f.u. with Nb at (1/3, 2/3, 1/4) and N at (0, 0, 1/2) and (0, 0, 1/4). For WC-NbN, the unit cell contains one f.u.
with Nb occupying (0, 0, 0) and N at (1/3, 2/3, 1/2). In the case of $\delta^\prime$-NbN, the unit cell contains two f.u. with Nb at (1/3, 2/3, 1/4) and N at (0, 0, 0). The four crystal structures are shown in Fig. 1.

\begin{figure}
\centering
\includegraphics[width=80mm]{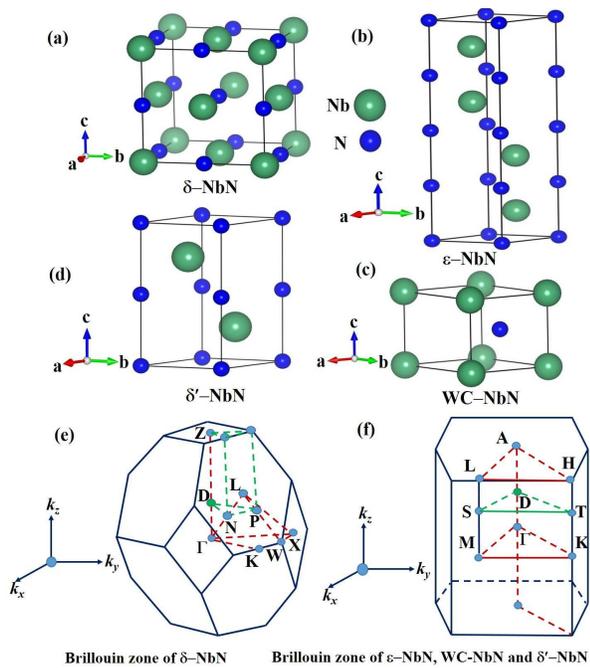}
\caption{Crystal structures of (a) $\delta$-NbN, (b) $\varepsilon$-NbN, (c) WC-NbN and (d) $\delta^\prime$-NbN.
The corresponding Brillouin zones of cubic $\delta$-NbN as well as hexagonal $\varepsilon$-NbN, WC-NbN and $\delta^\prime$-NbN
are schematically shown in (e) and (f), respectively. The N-D-P-N and S-D-T-S $k$-point paths shown in (e) and (f), 
respectively, are on the planes passing through the Dirac points D (see Fig. 4 below).
}
\end{figure}

{\it Ab initio} density functional theory (DFT) calculations are carried out with the generalized gradient approximation (GGA).\cite{PBE}
The electronic structure calculations are performed with the accurate projector-augmented wave
method \cite{GKresse, PEBlochl, Furthmuller} as implemented in the Vienna \textit{Ab initio} Simulation Package (VASP).
For the Brillouin zone integration with the tetrahedron method, $\Gamma$-centered $k$-point meshes
of 12 $\times$ 12 $\times$ 12, 12 $\times$ 12 $\times$ 4, 12 $\times$ 12 $\times$ 16 and 12 $\times$ 12 $\times$ 6 are used,
respectively, for $\delta$-NbN, $\varepsilon$-NbN, WC-NbN and $\delta^\prime$-NbN.
A large plane-wave cut-off energy of 500 eV is used throughout. The electronic density of states are calculated by
using denser $k$-point meshes of 16 $\times$ 16 $\times$ 16 for $\delta$-NbN, 18 $\times$ 18 $\times$ 6
for $\varepsilon$-NbN, 18 $\times$ 18 $\times$ 22 for WC-NbN and 18 $\times$ 18 $\times$ 9 for $\delta^\prime$-NbN.
A small total energy convergence criterion of 10$^{-6}$ eV is used for all the calculations.

The elastic constants of the NbN polytypes are determined by using the linear-response stress-strain method,
as implemented in the VASP code\cite{Page}.
For a crystal under a small strain ($\varepsilon_{kl}$), the corresponding stress ($\sigma_{ij}$) is
given by Hooke's law
\begin{equation}
\sigma_{ij} = C_{ijkl}\varepsilon_{kl},
\end{equation}
where $C_{ijkl}$ is the elastic tensor that comprises the elastic constants of the crystal.
When the symmetry of the crystal is taken into consideration, the total number of elastic constants can be reduced from 81.
In particular, a cubic structure has only three elastic constants of $C_{11}$, $C_{12}$ and $C_{44}$,
and a hexagonal structhre has five independent elastic constants of $C_{11}$, $C_{12}$, $C_{13}$, $C_{33}$ and $C_{44}$.~\cite{GY, Hill}
The bulk modulus $B$ and shear modulus $G$ are related to the elastic constants via $B = \frac{1}{3}(C_{11} + 2C_{12})$ and
$G = \frac{1}{5}(3C_{44} + C_{11} - C_{12})$ for cubic crystals. For hexagonal crystals, they are given by
$ B = \frac{2}{9} (C_{11} + C_{12} + 2C_{13} + \frac{1}{2}C_{33})$ and
$G = \frac{1}{30}(12C_{44} + 7C_{11} - 5C_{12} + 2C_{33} - 4C_{13})$.
The Young's modulus \textit{Y} is related to $B$ and \textit{G} by $Y = 9BG/(3B + G)$.

The strength of the electron-phonon coupling in a crystal is measured by the electron-phonon coupling constant ($\lambda$)
which can be extracted from the Eliashberg spectral function [$\alpha^2F(\omega)$] via~\cite{McMillan, AllenDynes}
\begin{equation}
\lambda = 2 \int \frac{\alpha^2F(\omega)}{\omega}d\omega.
\end{equation}
The Eliashberg spectral function is given by
\begin{equation}
\alpha^2F(\omega)=\frac{1}{2\pi N(\varepsilon_F)}\sum_{qj} \frac{\gamma_{qj}}{\omega_{qj}}\delta(\hbar\omega-\hbar\omega_{qj}),
\end{equation}
where $N(\varepsilon_F)$ is the electronic density of states at the Fermi level ($\varepsilon_F$), $\gamma_{qj}$ is
the phonon linewidth due to electron-phonon scattering, $\omega_{qj}$ is the phonon frequency of branch index $j$ at wave vector $q$.
Using the value of $\lambda$, one can estimate the superconducting transition temperature $T_c$
via McMillan-Allen-Dynes formula \cite{McMillan, AllenDynes}
\begin{equation}
T_c = \frac{\omega_{log}}{1.2} \textrm{exp}\Big[\frac{-1.04{(1+\lambda)}}{\lambda-\mu^*(1+0.62\lambda)}\Big],
\end{equation}
where $\omega_{log}$ is logarithmic average phonon frequency and $\mu^*$ is the averaged screened electron-electron interaction.

In the present study, the phonon dispersions, phonon density of states and electron-phonon interactions are computed
using the density functional perturbation theory \cite{SBaroni}, as implemented in the Quantum Espresso code.\cite{PGianozzi}
All the calculations are performed using the scalar-relativistic norm-conserving pseudopotentials. The plane wave cut-off
energy is set to 42 Ry for all the four structures of NbN. The electronic charge density is expanded up to 168 Ry.
A Gaussian broadening of 0.02 Ry is used for all the calculations except for $\delta$-NbN, where we consider
a range of values between 0.02 Ry to 0.18 Ry. The phonon calculations are performed with $q$-grids of 6 $\times$ 6 $\times$ 6,
6 $\times$ 6 $\times$ 3, 6 $\times$ 6 $\times$ 8 and 6 $\times$ 6 $\times$ 4 for $\delta$-NbN, $\varepsilon$-NbN, WC-NbN and $\delta^\prime$-NbN, respectively. The hole doping calculations are carried out by changing the total number of electrons in one unit cell and repeat the calculations for each doping concentration.

\begin{table}[ht]
\caption{Theoretical equilibrium lattice constants ($a, c, c/a$), volume ($V$) and total energy ($E_t$)
of $\delta$-NbN, $\varepsilon$-NbN, WC-NbN and  $\delta^\prime$-NbN compared with the experimental data\cite{GBrauer1,GBrauer,NTerao,Terao,Gold}.
}
\begin{tabular}{ccccccccc} \hline \hline
Phase          & $a$ (\AA)  & $c$ (\AA) & $c/a$ & V (\AA$^3$/f.u.) & $E_t$ (eV/f.u.)& \\ \hline
$\delta$-NbN   &  4.425 & 4.425 & 1.000 & 21.67 & -20.1228  &  \\
Expt\footnotemark[1]   & 4.391& 4.391 & 1.000 & 21.16 &   &   \\
$\varepsilon$-NbN & 2.974  & 11.332 & 3.810 & 21.70 & -20.5390  &    \\
 Expt\footnotemark[2]   & 2.96  & 11.27 & 3.807 & 21.38 &   &   \\
 WC-NbN & 2.952   &2.872  & 0.972 & 21.68 & -20.5049  &    \\
 Expt\footnotemark[3]              & 2.951 & 2.772 &0.939  & 20.91 &   & \\
 $\delta^\prime$-NbN & 2.981   &5.586  & 1.873 & 21.50 & -20.4705  &    \\
 Expt\footnotemark[4]      & 2.967 & 5.538 &1.866  & 21.11 &   & \\ \hline \hline
\end{tabular}\\
\footnotemark[1]{References [\onlinecite{GBrauer1,GBrauer}] (experiment);}
\footnotemark[2]{Reference [\onlinecite{Terao}] (experiment);}
\footnotemark[3]{Reference [\onlinecite{Gold}] (experiment);}
\footnotemark[4]{Reference [\onlinecite{NTerao}] (experiment).}
\end{table}

\section{RESULTS AND DISCUSSION}
\subsection{Mechanical properties}
As the first step, we determine theoretically the equilibrum lattice constants of $\delta$-NbN, $\varepsilon$-NbN, WC-NbN and $\delta^\prime$-NbN.
In Table I, we list the calculated lattice constants together with the available experimental data.~\cite{GBrauer1,GBrauer,NTerao,Terao,Gold}
The calculated lattice constants are found to be in good agreement (within 1\%) with the corresponding experimental data,
although the equilibrium volumes are slighty larger than the experimental ones, due to the fact that the GGA calculations
tend to overestimate the equilibrium lattice constants.\cite{JPPerdew} Table I also shows that $\varepsilon$-NbN
is the ground state structure while $\delta$-NbN, WC-NbN and $\delta^\prime$-NbN structures are, respectively, 0.416 eV/f.u., 0.034 eV/f.u. and 0.068 eV/f.u.
higher in total energy than $\varepsilon$-NbN.
Our obtained relative structural stabilities as well as lattice constants agree very well with the previous GGA calculations.\cite{Olifan}

\begin{table}[ht]
\caption{Calculated elastic constants ($C_{ij}$), bulk modulus ($B$), shear modulus ($G$) and Young's modulus ($Y$)
of $\delta$-NbN, $\varepsilon$-NbN, WC-NbN and $\delta^\prime$-NbN compared with the available experimental values.
The theoretical bulk modulii of cubic and hexagonal diamonds are also listed for comparison.
All these quantities are in units of GPa.}
\begin{tabular}{ccccccccccccccc} \hline \hline
 & $C_{11}$ & $C_{12}$ & $C_{13}$ & $C_{33}$ & $C_{44}$ & $B$ & $G$ & $Y$\\ \hline
$\delta$-NbN& 692 & 145.4 &  &  &65 & 327 & 148 & 385\\
 Expt\footnotemark[1]& 608  & 134   &  & & 117  &    292 &  165  &   \\
c-BN\footnotemark[2] & &  &  & & &    369 & &   \\
c-Diamond\footnotemark[3] & &  &  & & &    443 & &   \\
$\varepsilon$-NbN& 588 & 218 & 170 & 706 & 185  & 333 & 199 & 497 \\
 Expt\footnotemark[4] & &  &  & & &    373.2 & 200.5 & \\
WC-NbN &593 &235  &158  &812 &178  & 344 & 203 & 508\\
$\delta^\prime$-NbN &596 &228  &181  &642 &184  & 334 & 193 & 485\\
h-BN\footnotemark[5] & &  &  & & &    335 & &   \\
h-Diamond\footnotemark[6] & &  &  & & & 447 & &   \\ \hline \hline
\end{tabular}
\footnotemark[1]{Reference [\onlinecite{XJChen}] (experiment);}
\footnotemark[2]{Reference [\onlinecite{Knittle}] ({\it ab initio} calculation);}
\footnotemark[3]{Reference [\onlinecite{Klein}] ({\it ab initio} calculation);}
\footnotemark[4]{Reference [\onlinecite{YZou1}] (experiment);}
\footnotemark[5]{Reference [\onlinecite{hexaBN}] ({\it ab initio} calculation);}
\footnotemark[6]{Reference [\onlinecite{hexadiamond}] ({\it ab initio} calculation).}
\end{table}

Elastic constants are calculated for all the four structures of NbN, as tabulated in Table II.
All the elastic constants are found to be positive and follow the Born's \cite{Born} mechanical stability criteria,
i.e.,  $C_{11} > 0$, $C_{11} > C_{12}$, $C_{44} > 0$ for the cubic structure,
and $C_{11} > 0$, $C_{33} > 0$, $C_{11}-C_{12} > 0$, $C_{44} > 0$, and $(C_{11}+2C_{12})C_{33}-2C_{13}^2 > 0$
for the three hexagonal structures. Therefore, all the four structures are mechanically stable 
and hold the stability against specific deformations. This explains why all four NbN polytypes could
be prepared~\cite{GBrauer1,GBrauer,NTerao,Terao,Gold}, even although $\delta$-NbN, WC-NbN and $\delta^\prime$-NbN 
are metastable phases (Table I). The compressibility characteristics can be related
to the calculated elastic constants. According to high pressure experiments~\cite{YZou},
$\varepsilon$-NbN is more compressible along $a$-axis than $c$-axis. In consistence with this experimental
result, Table II shows that for $\varepsilon$-NbN, the value of $C_{33}$ is larger than $C_{11}$,
indicating that the material is harder to compress along $c$-axis.
The similar trend is also found in the other hexagonal structures WC-NbN and $\delta^\prime$-NbN. In the case of $\delta$-NbN,
the ordering of the calculated elastic constants of C$_{11}$ $>$ C$_{12}$ $>$ C$_{44}$ is also consistent
with the experimental one reported in Ref. [\onlinecite{XJChen}]. Our calculated elastic constants of $\delta$-NbN and WC-NbN
are in good agreement with previous GGA calculations. \cite{ZH,Jiang,DHolec,Raja,Meng}
However, there is no theoretical report on the GGA elastic constants of $\varepsilon$-NbN and $\delta^\prime$-NbN.

Table II indicates that all the four polytypes are hard materials with their bulk moduli being comparable to
that of super-hard cubic and hexagonal boron nitrides (c-BN and h-BN). The bulk modulus of $\delta$-NbN
is only about 10 \% lower than that of c-BN \cite{Knittle}, whereas for $\varepsilon$-NbN, WC-NbN and $\delta^\prime$-NbN,
the value is nearly the same as that of hexagonal boron nitride (h-BN)\cite{hexaBN}.
This indicates that the bonding in NbN polytypes is similar to that of boron nitride
which is primarily covalent in nature.\cite{hexaBN}
Furthermore, the bulk moduli of $\delta$-NbN, $\varepsilon$-NbN, WC-NbN and $\delta^\prime$-NbN are only lower by about 1/4
than those of cubic diamond \cite{Klein} and hexagonal diamond \cite{hexadiamond}, the hardest materials on Earth.
The reduction of about 1/4 of bulk modulus of $\delta$-NbN, $\varepsilon$-NbN, WC-NbN and $\delta^\prime$-NbN
could be attributed to the fact that the bonding in the diamonds involves three-dimensional network of atoms
whereas it is almost linearly distributed between Nb and N atoms in the NbN structures.
Young's modulus (\textit{Y}) is an important mechanical property of a crystalline material that specifies its stiffness.
The calculated \textit{Y} value of the three hexagonal structures of NbN is about 1/3 larger than that of $\delta$-NbN,
indicating the higher stiffness character of $\varepsilon$-NbN, WC-NbN and $\delta^\prime$-NbN. The ductility and brittleness characteristics
of a crystal can be analysed through Pugh's criteria. Accordingly, the $B/G$ ratio being greater than 1.75 indicates
ductile nature while being less than 1.75 resembles the brittle nature. The calculated $B/G$ values of 1.67
for $\varepsilon$-NbN, 1.69 for WC-NbN and 1.73 for $\delta^\prime$-NbN suggest that the hexagonal phases have brittle character. On the other hand,
the large $B/G$ value of 2.20 for $\delta$-NbN implies its ductile character.
Overall, the present study shows that $\varepsilon$-NbN is more brittle than WC-NbN, $\delta^\prime$-NbN and $\delta$-NbN.

\begin{figure}
\centering
\includegraphics[width=80mm]{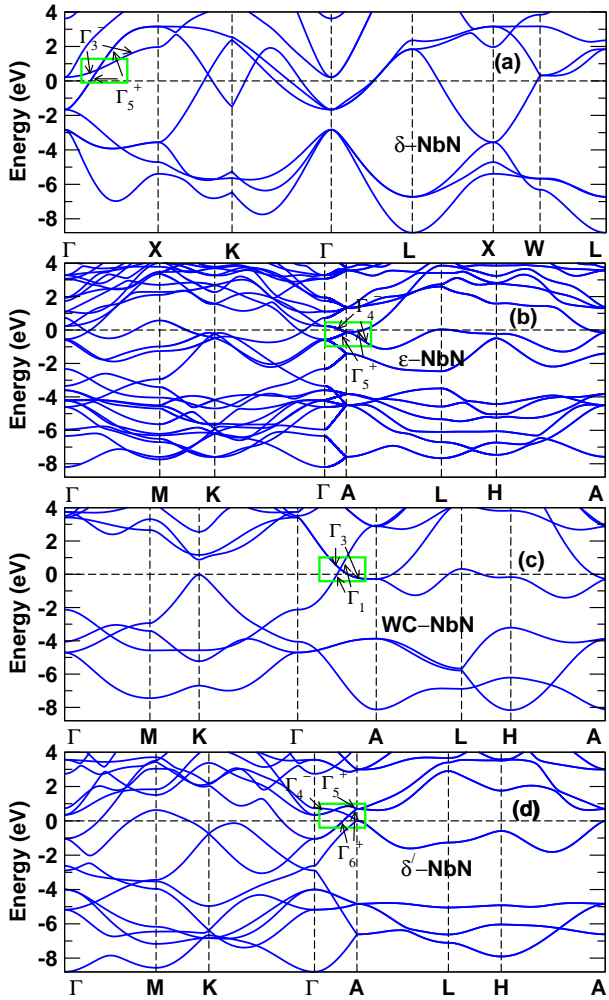}
\caption{Electronic band structures of (a) $\delta$-NbN (b) $\varepsilon$-NbN (c) WC-NbN and (d) $\delta^\prime$-NbN 
calculated without SOC. The green boxes indicate the band crossings discussed in the text, and the symmetries
of the crossing bands are labelled.}
\end{figure}

\subsection{Electronic band structure}
The study of electronic band structure is required for the detailed understanding of the physical properties
of the NbN polytypes. In Fig. 2, we display the band structures calculated without including the spin-orbit coupling (SOC).
The associated density of states (DOS) spectra are plotted in Fig. 3. Figures 2(a) and 2(c) show that the band structures of $\delta$-NbN
and WC-NbN consist of three filled low lying valence bands and partially occupied conduction bands above them.
It can be seen from Fig. 3(a) that the three valence bands are strongly Nb $d$ and N $p$ orbital hybridized bands
while the conduction bands are made of mainly Nb $d$ orbital. The band structures of $\varepsilon$-NbN 
and $\delta^\prime$-NbN are more complicated simply because the number of atoms per unit cell in $\varepsilon$-NbN 
is four times and in $\delta^\prime$-NbN two times that of $\delta$-NbN and WC-NbN. Nonetheless, 
we can see from Figs. 2(b) and 2(d) that there are now 12 and 6 low lying valence bands with a strong mixture of
Nb $d$ and N $p$ orbitals and above these lower conduction bands made of mainly Nb $d$ orbital 
[see Fig. 3(b) and Fig. 3(d)] in $\varepsilon$-NbN and $\delta^\prime$-NbN, respectively.

Figure 3(a) shows that in $\delta$-NbN, the valence bands below -4.0 eV are strongly Nb $d$ and N $p$ orbital 
hybridized bands with almost equal weights of Nb $d$ and N $p$ orbitals, indicating strongly covalent bonding nature.
Their DOS spectrum features a prominant peak near -6.0 eV.
Above -4.0 eV, the bands consist of mainly Nb $d$ orbital and their DOS increases almost linearly
with energy from -4.0 eV to 1.0 eV.
This results in a rather large DOS at the Fermi level (0.90 states/eV/f.u.).

\begin{figure}
\centering
\includegraphics[width=80mm]{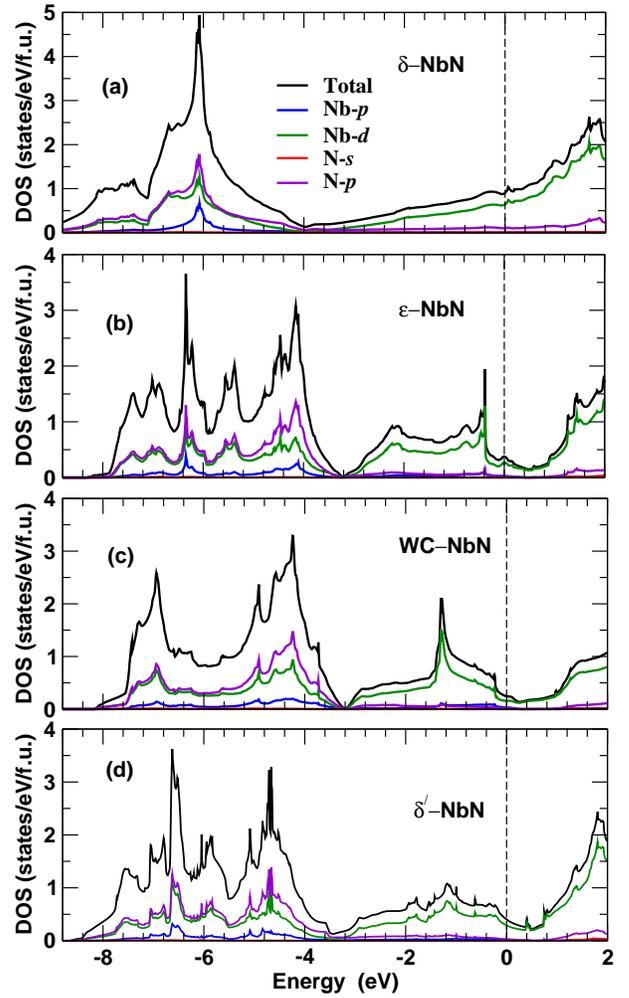}
\caption{Total and orbital-decomposed Density of states (DOS) of (a) $\delta$-NbN, (b) $\varepsilon$-NbN, 
(c) WC-NbN and (d) $\delta^\prime$-NbN.}
\end{figure}

Similarly, in $\varepsilon$-NbN, WC-NbN and $\delta^\prime$-NbN the valence bands below -3.2 eV also consist of 
strongly Nb $d$ and N $p$ orbital hybridized bands with nearly equal weights of Nb $d$ and N $p$ orbitals 
[see Figs 3(b), 3(c) and 3(d)]. This strongly covalent bonding between Nb and N
could explain why all four polytypes are hard. Also, the bands above -3.2 eV in $\varepsilon$-NbN, WC-NbN and $\delta^\prime$-NbN are
again composed of mainly Nb $d$ orbital. However, the DOS spectrum does not increase monotonically with energy.
Instead, there is a pseudogap centered at $\sim$1.0 eV above the Fermi level.
This gives rise to a smaller DOS at the Fermi level of 0.33 states/eV/f.u.
for $\varepsilon$-NbN, 0.24 states/eV/f.u. for WC-NbN and 0.39 states/eV/f.u. for $\delta^\prime$-NbN.
This also leads to the fact that $\varepsilon$-NbN, WC-NbN and $\delta^\prime$-NbN are more stable than $\delta$-NbN (Table I) and
that $\varepsilon$-NbN, WC-NbN and $\delta^\prime$-NbN possess somewhat superior mechanical properties compared to cubic $\delta$-NbN.

\begin{figure*}
\centering
\includegraphics[width=160mm]{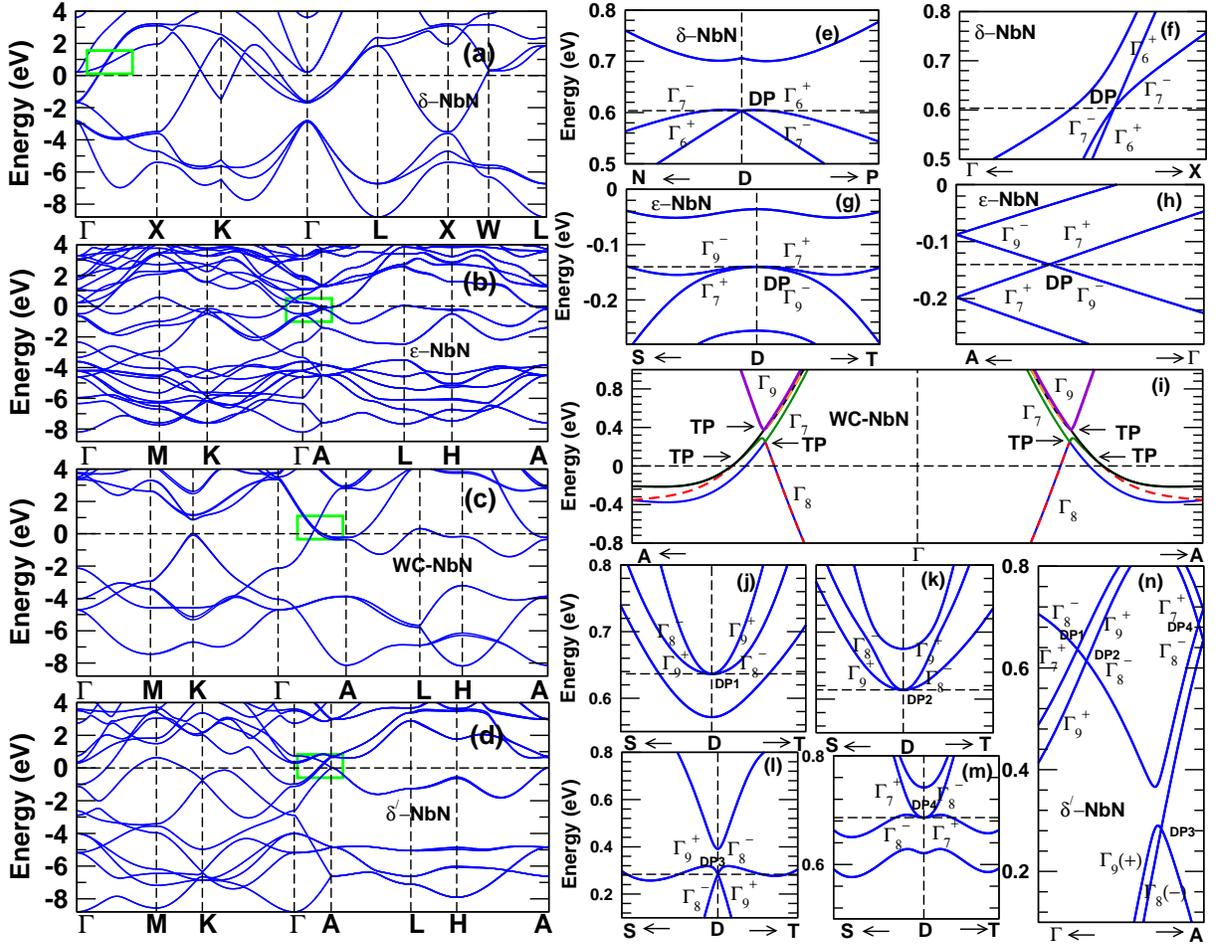}
\caption{Relativistic electronic band structures of (a) $\delta$-NbN, (b) $\varepsilon$-NbN, (c) WC-NbN and (d) for $\delta^\prime$-NbN.
Panels (e)+(f), (g)+(h), (i), (j)+(k)+(l)+(m)+(n) are the zoom-in plots of the band crossings in the green boxes in panels (a), (b) (c) 
and (d) respectively. DP in panels (e), (f), (g), (h), (j), (k), (l), and (m) and (n), and TP in panel (i) denote, respectively, 
Dirac point and triple nodal point. The energy bands together with their symmetries 
along the in-plane paths N-D-P and S-D-T through the DPs (see Fig. 1)
are shown, respectively, for $\delta$-NbN (e) and for both $\varepsilon$-NbN (g) and $\delta^\prime$-NbN (j),(k), (l), (m).
}
\end{figure*}

Interestingly, Fig. 2 shows that there are quite a few band crossings in the vicinity of
the Fermi level in the band structures of the four NbN polytypes.
In the case of $\delta$-NbN, the band crossing along the $\Gamma$-X path belongs to little point group $C_{4v}$ and 
the two crossing bands have distinct irreducible representations (IRs) and parities of $\Gamma_3^-$ and $\Gamma_5^+$
[see the green box in Fig. 2(a)]. Consequently, the two bands cannot mix and thus the band crossing is 
protected by the $\mathcal {C}_{4}$ rotational symmetry. For hexagonal $\varepsilon$-NbN, 
the linear band crossing point is located just below the Fermi level at $k$-point A. The two crossing bands belong to 
IRs $\Gamma_4^-$ and $\Gamma_5^+$ of the $C_{6v}$ point group [see Fig. 2(b)] and it is thus protected by the 
three-fold $\mathcal {C}_{3z}$ rotational symmetry. There also exists a band crossing along $\Gamma$-A 
with IRs of $\Gamma_1$ and $\Gamma_3$ 
of little point group $C_{3v}$ in WC-NbN. This band crossing is protected by the $\mathcal {C}_{3z}$ rotational symmetry 
and also mirror symmetries $\mathcal{M}_y$ and $\mathcal{M}_z$. In $\delta^\prime$-NbN, three band crossings occur 
in the vicinity of the Fermi level along $\Gamma$-A. The three bands involved in the band crossings 
belong to IRs $\Gamma_4^-$, $\Gamma_5^+$ and $\Gamma_6^+$ 
of the $C_{6v}$ little group, as shown in Fig. 2(d). In short, all the bands involved in the band crossings 
shown in the green boxes in Fig. 2 for the four NbN-polytypes, have different IRs and hence the crossings are unavoidable. 
This suggests that these NbN polytypes could be topological metals that would host three-dimensional (3D) 
Dirac fermions~\cite{Fu07,Young12,Wang12,Yang14,Yan17} or other emergent fermions~\cite{Bian,Lv17}.

To be topological metals, these band crossings should remain ungapped when the SOC is included.
Therefore, we calculate the fully relativistic band structures of these NbN polytypes which
are displayed in Fig. 4. Remarkably, the band crossings along the rotational axes
(see those surrounded by the green boxes in Figs. 2 and 4), i.e., the band crossing along the $\Gamma$-X line 
in $\delta$-NbN [Figs. 4(a) and 4(f)] as well as the crossing along the $\Gamma$-A line 
in $\varepsilon$-NbN [Figs. 4(b) and 4(h)], WC-NbN [Figs. 4(c) and 4(i)] and also in $\delta^\prime$-NbN [Figs. 4(d) and 4(n)],
remain intact when the SOC is included in the band structure calculations. To verify that these band crossings 
are unavoidable, we determine the IR for each crossing band in all the four polytypes. Interestingly, 
the two crossing bands in $\delta$-NbN have IRs $\Gamma_7^-$ and $\Gamma_6^+$ of the $C_{4v}$ double point group, 
as shown in Figs. 4(e) and 4(f), and it is thus protected by the $\mathcal {C}_{4}$ rotational symmetry. 
For $\varepsilon$-NbN, the two crossing bands shown in Figs. 4(g) and 4(h) have different IRs of $\Gamma_7^+$ 
and $\Gamma_9^-$ of the $C_{6v}$ double point group and hence it is protected by the three-fold $C_{3z}$ rotational symmetry. 
In the case of WC-NbN, due to the lack of the spatial inversion symmetry the crossing bands split 
into non-degenerate bands of IR $\Gamma_7$ and doubly degenerate bands of IRs  $\Gamma_8$ and $\Gamma_9$,
thus resulting in triply degenerate crossing points protected by the $\mathcal {C}_{3z}$ rotational symmetry
[see Figs. 4(c) and 4(i)]. The band crossings in $\delta^\prime$-NbN belong to IRs $\Gamma_9^+$ and $\Gamma_8^-$ [Figs. 4(j) and 4(n)], 
$\Gamma_7^+$ and $\Gamma_8^-$ [Figs. 4(k) and 4(n)], $\Gamma_9^+$ and $\Gamma_8^-$ [Figs. 4(j) and 4(n)] 
and $\Gamma_7^+$ and $\Gamma_8^-$ [Figs. 4(m) and 4(n)]. Consequently, these band crossings are protected 
by the $\mathcal {C}_{3}$ rotational symmetry. Furthermore, to see these topological nodal points more clearly,
we also display the energy bands near the crossing points (D) along the N-D-P path for $\delta$-NbN [Fig. 4(e)] and along
the S-D-T path for both $\varepsilon$-NbN [Figs. 4(g)] and $\delta^\prime$-NbN [Figs. 4(j), 4(k), 4(l) and 4(m)] in the in-planes going 
through the crossing (Dirac) point D (see Fig. 1). This demonstrates that all four NbN polytypes are topological metals.
All the other band crossings in Fig. 2 become gapped when the SOC is included (see Fig. 4).
This could be expected because certain crystalline symmetries such as threefold and fourfold rotations
are needed to protect these 3D band crossing points.~\cite{Young12}

Remarkably, these band crossings in the four different structures of NbN belong, respectively,
to three different kinds of topological nodal points, namely, conventional (i.e., type-I) and type-II
Dirac points (DPs)~\cite{Fu07,Young12,Wang12,Yang14,Yan17} as well as triply degenerate nodal points (TPs)~\cite{Bian,Lv17}.
$\delta$-NbN, $\varepsilon$-NbN and $\delta^\prime$-NbN have both time-reversal ($\mathcal{T}$) symmetry and spatial inversion ($\mathcal{P}$)
symmetry, and thus each of their energy bands is twofold degenerate. Consequently, the band crossings in these structures
are fourfold DPs.~\cite{Fu07,Young12,Wang12,Yang14,Yan17} In $\varepsilon$-NbN and $\delta^\prime$-NbN, 
the DPs [see Figs. 4(b), 4(g), 4(h) and Figs. 4(d), 4(j), 4(k), 4(l), 4(m), 4(n)]
are the conventional one as the slopes of the crossing bands have opposite signs.~\cite{Fu07,Young12,Wang12,Yang14}
In contrast, the DPs in $\delta$-NbN [see Figs. 4(a), 4(e) and 4(f)]
are rare type-II Dirac points since the slopes of the crossing bands have the same sign.~\cite{Yan17}
Moreover, the band crossings in WC-NbN are exotic triply degenerate nodal points~\cite{Bian,Lv17}
that may host emergent fermions which are absent in high-energy physics.
WC-NbN has broken $\mathcal{P}$ symmetry and consequently its energy bands may
split into nondegenerate ones away from the $\mathcal{T}$ symmetric $k$-points
in the Brillouin zone. Therefore, the state degeneracies of band crossings in WC-NbN
could be an odd number such as three in the present case.

\begin{figure}
\centering
\includegraphics[width=80mm]{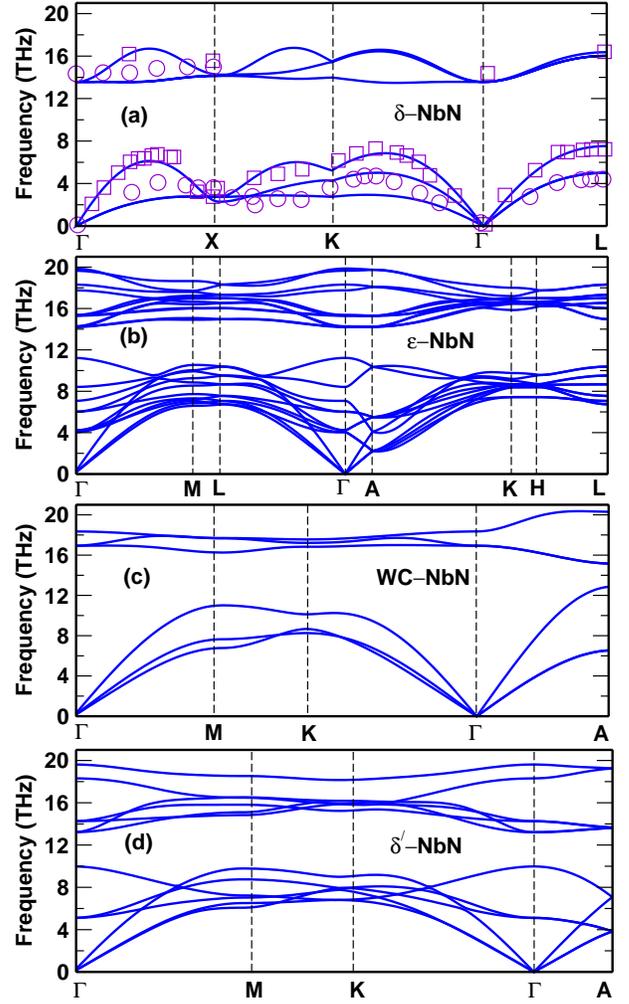}
\caption{Phonon dispersion relations of (a) $\delta$-NbN, (b) $\varepsilon$-NbN, (c) WC-NbN and (d) $\delta^\prime$-NbN.
The electronic smearing width ($\sigma$) used in the phonon calculations is $\sigma = 0.18$
Ry for (a) and $\sigma = 0.02$ Ry for (b), (c) and (d).
In (a), neutron scattering data \cite{Christ} are also plotted in open symbols for comparison.}
\end{figure}

\subsection{Phonon dispersion relations}
Now we turn our attention towards the phonon dispersion spectra of NbN ploytypes. The calculated phonon
dispersion relations for all the NbN structures are displayed in Fig. 5. The associated phonon DOSs
are plotted in Fig. 6.  Since $\delta$-NbN and WC-NbN have two atoms per unit cell,
their phonon dispersion relations have six branches with three acoustic and three optical modes
[see Figs. 5(a) and 5(c)]. There is a gap separating the optical bands from the acoustic
bands [Figs. 5(a) and 5(c) as well as Figs. 6(a) and 6(c)].
The gap arises because of the large mass difference between Nb and N atoms.
The acoustic bands come predominantly from the vibrations of heavier Nb atoms while the optical branches
are mainly due to lighter N atoms. The acoustic bands are rather dispersive while
the optical bands are rather narrow. In $\varepsilon$-NbN ($\delta^\prime$-NbN),
the unit cell has 8 (4) atoms and thus there are 24 (12) phonon branches
[Figs. 5(b) and (5(d))]. Out of the 24 (12) phonon branches three are acoustic and remaining 21 (9) are optical.
There is no gap separating the three acoustic bands from the optical bands.
Nonetheless, the 24 (12) bands can be divided into two groups with a gap separating them,
namely, 12 (6) low lying dispersive bands dominated by the Nb vibrations and 12 (6) high lying narrower bands
arising from the N vibrations [Figs. 5(b), (5(d)) and 6(b), (6(d))].

Figure 5 show that all the phonon frequencies of the four NbN polytypes are positive and this
means that they are all dynamically stable. Nonetheless, we should note that in $\delta$-NbN,
all the calculated phonon frequencies are positive only when an abnormally large value of the electronic
band smearing width ($\sigma \ge 0.15$ Ry) is used. If an ordinary value of $\sigma$, e.g., $\sigma = 0.02$ Ry,
is used, the calculated phonon frequencies of the acoustic branches in the vicinity of the X-point and K-point
become imaginary (i.e., the phonon frequency squares become negative), as shown in Fig. 7(a) in the Appendix.
It is known that the presence of imaginary phonon frequencies in a crystalline material
indicates that its structure would become (dynamically) unstable. In other words,
$\delta$-structure of pure NbN would be unstable. The presence of imaginary phonon frequencies near the X-point
have been found before in previous GGA phonon frequency calculations.\cite{EIIsaev, Olifan, Jha}
In fact, this appears to be a common feature found for the superconducting transition metal
nitrides and carbides with NaCl structure.\cite{EIIsaev1,Olifan, Christ}
Indeed, experimentally, $\delta$-NbN phase could only be prepared with a small N deficiency
($x$) at high temperatures\cite{Christ,XJChen}. In $\delta$-NbN$_{1-x}$ with a small number of N vacancies
(V$_N$) ($x << 1.0$), N atoms and vacancies on the N sublattice are randomly distributed. The main
effect of the disorder due to this random distribution of N and V$_N$
on electronic band energies is a larger band smearing.\cite{Olifan} Therefore, to a first-order approximation,
the effect of the disorder due to nitrogen vacancies in NbN$_{1-x}$ could be taken into account by
using a large electronic smearing width in the theoretical calculations.\cite{Olifan}
Indeed, all the phonon frequencies of $\delta$-NbN shown in Fig. 5(a) calculated with $\sigma = 0.18$ Ry
are now positive. Moreover, they are in good agreement with the available neutron scattering
experiments\cite{Christ}. In particular, the calculated longitudinal acoustic phonon branch
agrees very well with the experimental one, although small discrepancies between the calculation
and experiment could be found for the transverse acoustic phonons [Fig. 5(a)].
Note that the neutron scattering experiments were carried out on $\delta$-NbN$_{0.93}$.
Also note that our calculated phonon dispersion relations of $\delta$-NbN are
in much better agreement with the neutron scattering data than that reported previously in Ref. [\onlinecite{Olifan}].
To further investigate the effect of the N deficiency we also calculate the phonon dispersion within the
virtual crystal approximation, i.e., the small N deficiency ($x$) in NbN$_{1-x}$ is simulated by a small reduction ($7x$) 
in the number of valence electrons. For example, to simulate $\delta$-NbN$_{0.93}$, we would reduce
the number of valence electrons by $\sim$0.5 e/f.u. We find that when the N deficiency $x$ in $\delta$-NbN$_{1-x}$ 
becomes more than 0.05, the imaginary phonon band disappears, indicating that the structure becomes dynamically stable. 
When the $x$ is further increased to $\sim$0.07, the soft acoustic phonon mode at the X point
becomes a normal phonon mode as shown in Fig. 7(b). Therefore, we conclude that the small 
N deficiency would indeed stabilize the $\delta$-NbN structure.

\begin{figure}
\centering
\includegraphics[width=80mm]{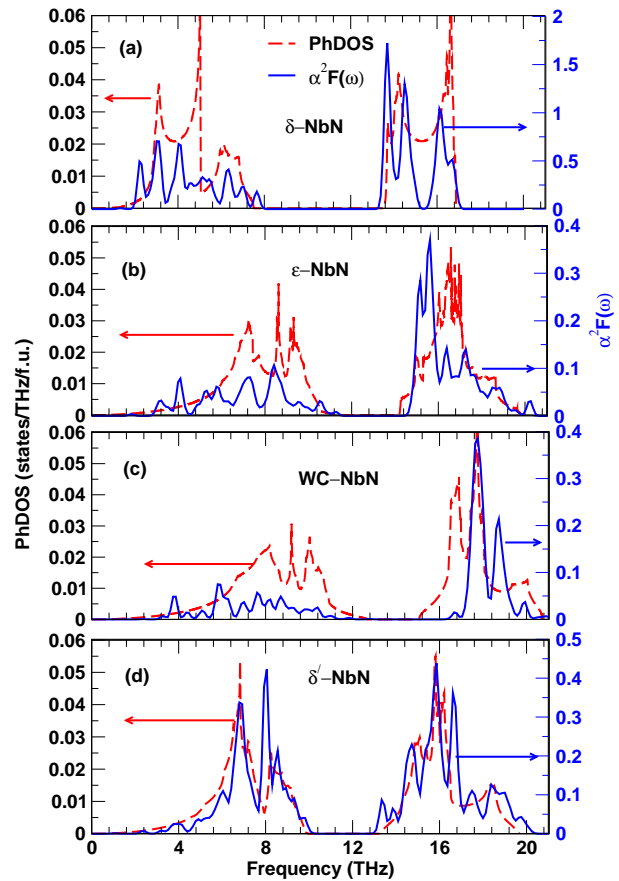}
\caption{Phonon density of states (PhDOS) and Eliashberg function [$\alpha^2F(\omega)$] of
(a) $\delta$-NbN, (b) $\varepsilon$-NbN (c) WC-NbN and (d) $\delta^\prime$-NbN.}
\end{figure}

On the other hand, all the phonon dispersion spectra of hexagonal $\varepsilon$-NbN, WC-NbN and $\delta^\prime$-NbN calculated
using different smearing widths of either $\sigma = 0.02$ Ry (Fig. 5) or $\sigma = 0.10$ Ry (Fig. 8)
are positive, implying that they are all dynamically stable.
Furthermore, a comparison of Fig. 5(b), 5(c) and 5(d), respectively, with Figs. 8(a), 8(b) and 8(c)
shows that the electronic band smearing width used has almost no effect on the calculated phonon dispersion relations. 
To further study the effect of the possible N deficiency on the phonon dispersion of hexagonal structures of NbN, 
we also calculate the phonon dispersion of $\varepsilon$-NbN$_{1-x}$  with a small $x$ of $\sim$0.02 within
the virtual crystal approximation. Figure 7(c) shows clearly that the small N deficiency has no effect 
on the phonon dispersion of $\varepsilon$-NbN.
Our calculated phonon dispersion relations of WC-NbN are in good agreement with that of
the earlier GGA calculation reported in Ref. [\onlinecite{EIIsaev1}] but differ from that
reported in Ref. [\onlinecite{Olifan}].
In particular, there is no visible splitting of the transverse and longitudinal optical phonon modes
at the $\Gamma$ in the phonon dispersion relations reported in Ref. [\onlinecite{Olifan}], which
is incorrect. However, no experimental measurements nor theoretical calculations
on the phonon dispersion relations of $\varepsilon$-NbN have been reported.

\subsection{Eliashberg function and superconductivity}
The main focus of the present study is to study the strength of the electron-phonon coupling
in all four NbN polytypes, which is given by an integral of Eliashberg function $\alpha^2F(\omega)$ times $1/\omega$
over phonon frequency [Eq. (2)]. Table III lists the calculated electron-phonon coupling constants ($\lambda$).
Figure 6 displays the calculated Eliashberg functions $\alpha^2F(\omega)$ along with the phonon DOSs.
Overall, each $\alpha^2F(\omega)$ function follows its corresponding phonon DOS spectrum.
For example, like the phonon DOS spectra, the $\alpha^2F(\omega)$ spectra
of all the four structures show a gap between the lower Nb-dominant and upper N-dominant phonon regions.
Interestingly, the gap as well as the centroids of the two phonon regions move up in energy
as one goes from $\delta$-NbN to $\delta^\prime$-NbN to $\varepsilon$-NbN and finally to WC-NbN (see Fig. 6).
This results in the fact that $\delta$-NbN has the smallest logarithmic average phonon frequency
$\omega_{log}$ while WC-NbN has the largest $\omega_{log}$, as shown in Table III.
Also, the $\alpha^2F(\omega)$ is considerably larger in the upper region
than in the lower region. Note that the two phonon regions have an equal number of phonon bands
and thus can accommodate an equal number of phonons, i. e., their areas
under the phonon DOS curves should be the same. Consequently, the phonon DOS and hence $\alpha^2F(\omega)$
are larger in the upper region than in the lower region because the upper region is narrower
than the lower region (Fig. 6). Remarkably, Fig. 6 show that the magnitude of the $\alpha^2F(\omega)$ spectrum
in $\delta$-NbN is significantly larger than that in $\varepsilon$-NbN, WC-NbN and $\delta^\prime$-NbN, 
thus implying a much stronger electron-phonon coupling in $\delta$-NbN. Note that the large peaks 
in $\alpha^2F(\omega)$ at around 14 THz for $\delta$-NbN results from the coupling of the transverse optical modes 
due to N atoms to the electrons. This, together with the lower centroids of the two phonon regions in $\delta$-NbN, 
suggests that $\delta$-NbN would have the highest $\lambda$ value. Indeed, Table III shows that $\delta$-NbN 
has a value of $\lambda$ (0.98) much larger than that of $\delta^\prime$-NbN (0.17), $\varepsilon$-NbN (0.16) 
and WC-NbN (0.11). Interestingly, we find that the major contribution to the electron-phonon coupling
constant $\lambda$ in the NbN polytypes comes from the Nb vibration dominated phonon modes 
in the low frequency region. For example, the contribution of these phonon modes to 
the $\lambda$ in $\delta$-NbN is nearly 80 \%. Therefore, it is easy to see from Eq. (2)
why  $\delta$-NbN has the largest $\lambda$ value among the four NbN polytypes. 

\begin{table}[ht]
\caption{Calculated electron-phonon coupling constant ($\lambda$), logarithmic average phonon frequency
($\omega_{log}$), Debye temperature ($\Theta_D$), density of states at the Fermi level $N(\varepsilon_F)$
and superconducting transition temperature ($T_c$) of $\delta$-NbN, $\varepsilon$-NbN, 
WC-NbN and $\delta^\prime$-NbN. The smearing parameter ($\sigma$) used in these calculations
is set to $\sigma = 0.18$ Ry, 0.02 Ry for $\delta$-NbN and to $\sigma = 0.02$ Ry, 0.10 Ry and 0.18 Ry 
for $\varepsilon$-NbN, WC-NbN and $\delta^\prime$-NbN.
The screened Coulomb interaction $\mu^*$ entering Eq. (4)
is set to 0.10. Available experimental $T_c$ values are also listed for comparison.}
\begin{ruledtabular}
\begin{tabular}{ccccccc}
Structure      &$\sigma$& $\lambda$ & $\omega_{log}$ & $\Theta_D$ & $N(\varepsilon_F)$ & $T_c$  \\
               & &           &            (K) &  (K)       & (states/eV/f.u.) & (K)  \\ \hline
$\delta$-NbN   &0.18&  0.98     & 269  & 637   & 0.883    & 18.26    \\
$\delta$-NbN$_{0.93}$  &0.02&  1.07     & 271  &    & 0.842    &  20.86   \\
Expt           &&  &                &     &    & 17.3\footnotemark[1]    \\
$\varepsilon$-NbN & 0.02& 0.16      & 398  & 734   & 0.331    & 0.00      \\
           &0.10&  0.27    & 456  &   & 0.339    &  0.08  \\
           &0.18& 0.36     & 472 &    & 0.341    &  1.08   \\
$\varepsilon$-NbN$_{0.97}$ & 0.02& 0.36    & 396  &   & 1.469    & 0.92      \\
Expt           & &  &  &  & & $<1.77$\footnotemark[2],11.6\footnotemark[3]  \\
 WC-NbN      & 0.02& 0.11       & 639  & 740 & 0.244   &  0.00 \\
           &0.10&   0.22   & 479  &   &   0.248  & 0.00   \\
           &0.18& 0.34     & 504 &    &  0.255   &  0.89   \\
$\delta^\prime$-NbN  &0.02    & 0.17      & 430  & 715 & 0.394   &  0.00 \\
                     &0.10&  0.35    & 483  &   &  0.397   & 0.90   \\
                     &0.18&  0.43    & 496 &    &0.411     &  2.91
\end{tabular}
\end{ruledtabular}
\footnotemark[1]{Reference [\onlinecite{Ralls}] (experiment);}
\footnotemark[2]{Reference [\onlinecite{Oya}] (experiment);}
\footnotemark[3]{Reference [\onlinecite{YZou1}] (experiment).}
\end{table}

The trend of the calculated $\lambda$ values
could also be understood in terms of the expression~\cite{McMillan}
$\lambda = [N(\varepsilon_F)/<\omega^2>]\sum_i(<I^2>_i/M_i)$ where $M_i$ is the atomic mass of atom $i$
and $<I^2>_i$ is the square of the electron-phonon coupling matrix element averaged over the
Fermi surface. Also, $<\omega^2> \approx 0.5 \Theta_D^2$ where the Debye temperature $\Theta_D$
could be related to the elastic constants\cite{Liu17}. Using the calculated elastic constants (Table II)
we estimate the $\Theta_D$ for all four NbN polytypes, as listed in Table III. Therefore,
it is clear from Table III that $\delta$-NbN has the largest $\lambda$ value
because it has the largest $N(\varepsilon_F)$ (Table III) and the smallest $\Theta_D$
(hence the smallest $<\omega^2>$).

By using the calculated electron-phonon coupling constant $\lambda$, we estimate the superconducting
transition temperature $T_c$ for all four NbN polytypes with Allen-Dynes formula [Eq. (4)] (see Table III). 
Here the screened electron-electron repulsion $\mu^*$ is treated as
an empirical parameter and is set to 0.10.~\cite{AllenDynes}
For $\delta$-NbN, the calculated $T_c$ is 18.2 K, being in good agreement with the experimental
value of 17.3 K \cite{Ralls}. No experimental finding of the superconductivity in
WC-NbN and $\delta^\prime$-NbN has been reported, and this seems to agree with our prediction 
of $T_c = 0$ K for pure WC-NbN and $\delta^\prime$-NbN with small smearing width $\sigma = 0.02$ Ry (Table III).
Oya and Onoders reported in 1973 that $\varepsilon$-NbN did not
exhibit superconductivity down to 1.77 K.\cite{Oya} This experimental result
is in agreement with our prediction of zero $T_c$ value for pure $\varepsilon$-NbN (Table III).
However, Zou {\it et al.} \cite{YZou} recently found two superconducting transitions at 17.5 K and 11.6 K
in their polycrystalline samples of mixed $\varepsilon$-NbN and $\delta$-NbN phases.
They attributed the superconducting transitions at 11.6 K and 17.5 K to the $\varepsilon$-NbN and $\delta$-NbN phases, respectively.
This appears to be in contradiction with the earlier experiment by Oya and Onoders \cite{Oya}
and also with the present calculation (Table III).

In order to ensure that our theoretical results are converged with respect to the
computational parameters used, we further perform the calculations
with a denser $q$-grid of 8 $\times$ 8 $\times$ 8 for $\delta$-NbN, 8 $\times$ 8 $\times$ 2 for $\varepsilon$-NbN, 
8 $\times$ 8 $\times$ 10 for WC-NbN and 8 $\times$ 8 $\times$ 6 for $\delta^\prime$-NbN. The calculated phonon DOS,
$\alpha^2F(\omega)$ and $\lambda$ etc remain almost the same.
Consequently, the calculated $T_c$ is still zero for pure $\varepsilon$-NbN, WC-NbN and $\delta^\prime$-NbN 
while the $T_c$ for $\delta$-NbN increases slightly to 19.1 K.
For transition metals and their binary compounds, the screened electron-electron repulsion $\mu^*$
usually ranges from 0.09 to 0.15.~\cite{AllenDynes}
We also calculate $T_c$ using $\mu^* = 0.13$ but find that the $T_c$ remains zero for $\varepsilon$-NbN, 
WC-NbN and $\delta^\prime$-NbN while the $T_c$ for $\delta$-NbN gets reduced slightly to 16.8 K.

Given the fact that as in $\delta$-NbN, there could be some N vacancies in $\varepsilon$-NbN, WC-NbN and $\delta^\prime$-NbN samples,
we also carry out further calculations using the electronic smearing widths of larger than 0.02 Ry.
Table III indicate that for $\sigma = 0.18$ Ry, the $T_c$ for $\varepsilon$-NbN, WC-NbN and $\delta^\prime$-NbN
become nonzero but small (1.0$\sim$3.0 K), due to substantially enhanced $\lambda$ values.
Since a larger smearing width has almost no effect on the phonon dispersion relations
in $\varepsilon$-NbN and WC-NbN (Fig. 5 and Fig. 8), the enhanced $\lambda$
could be attributed to the increased $N(\varepsilon_F)$. Figure 4 shows that as mentioned before,
the Fermi level sits on the slope of the lower energy side of the pseudogap. Consequently, when a much larger $\sigma$
value is used, the peak in the DOS spectrum just below the Fermi level becomes considerably broadened
and thus transfers some weight to the Fermi level, leading to an increased $N(\varepsilon_F)$. 
To further explore the consequences of small N deficiency, we also calculate the superconducting properties 
of $\delta$-NbN$_{0.93}$ and $\varepsilon$-NbN$_{0.97}$ within the virtual crystal approximation. 
As expected, Table III shows that the DOS of $\varepsilon$-NbN$_{0.97}$ at the Fermi level increases 
by about 4 times compared to the pure $\varepsilon$-NbN case [Fig. 3(b)]. This results in an enhanced $\lambda$,
thus leading to a $T_c$ of 0.92 K. Based on these results, we may conclude that if there were the superconductivity in 
hexagonal NbN polytypes, the superconducting transition temperature would be smaller than $\sim$ 1.0 (3.0) K 
in $\varepsilon$-NbN and WC-NbN ($\delta^\prime$-NbN). To clarify this important issue, 
we believe that further experiments on the single-phase samples of $\varepsilon$-NbN would be helpful.

\section{CONCLUSION}
Summarizing, we have investigated the mechanical properties, electronic structure, lattice dynamics,
electron-phonon interactions and superconductivity in all four NbN polytypes ($\delta$-NbN, $\varepsilon$-NbN, 
WC-NbN and $\delta^\prime$-NbN) by performing systematic {\it ab initio} DFT-GGA calculations.
The calculated total energy and elastic constants (Tables I and II) reveal that $\varepsilon$-NbN is the ground state structure
but $\delta$-NbN, WC-NbN and $\delta^\prime$-NbN are also mechanically stable, thus explaining the fact that all four NbN polytypes
have been reported. These results also indicate that all four polytypes are hard materials with
their bulk modulii being comparable to that of cubic and hexagonal BN~\cite{Knittle,hexaBN}.
In fact, their bulk modulii (Table II) are only about 1/4 smaller than that of cubic and hexagonal
diamond~\cite{Klein,hexadiamond}, the hardest materials on Earth.
The calculated electronic band structures (Figs. 2 and 3) show that all four polytypes
are metallic with the energy bands in the vicinity of the Fermi level ($E_F$) being dominated by Nb \textit{d}-orbitals.
Nonetheless, the lower part of the valence band manifold is of strongly covalent bonding with nearly equal weights of
Nb $d$ and N $p$ orbitals, thus resulting in large bulk and Young's modulii.

The calculated phonon dispersion relations (Fig. 5) can be divided into two groups separated by a band gap,
namely, low frequency heavier Nb vibration-dominated one and high frequency lighter N vibration-dominated one.
The calculated phonon dispersion relations of $\delta$-NbN are in excellent agreement with the available
neutron scattering experiments~\cite{Christ}.
Interestingly, the calculated phonon DOSs (Fig. 6) reveal that the centroids of the two groups and hence
the Debye temperature (Table III) go up as one moves from $\delta$-NbN to $\delta^\prime$-NbN to $\varepsilon$-NbN and then to WC-NbN.
The calculated Eliashberg functions follow the same trend and thus give rise to the
largest electron-phonon coupling constant of $\lambda = 0.98$ in $\delta$-NbN. $\delta^\prime$-NbN has value of $\lambda = 0.17$,
$\varepsilon$-NbN has a much smaller $\lambda$ of 0.16 and WC-NbN has the smallest $\lambda$
of 0.11 (Table III). This trend of the $\lambda$ values can be attributed to the
trend of the DOS at the Fermi level $N(\varepsilon_F)$, viz,
$N(\varepsilon_F)^{\delta} > N(\varepsilon_F)^{\delta^\prime} > N(\varepsilon_F)^{\varepsilon} > N(\varepsilon_F)^{WC}$,
and also that of Debye temperature $\Theta_D$, i.e., $\Theta_D^{\delta} < \Theta_D^{\delta^\prime} < \Theta_D^{\varepsilon} < \Theta_D^{WC}$,
of the four NbN polytypes.
The estimated superconducting transition temperature $T_c$ of 18.2 K of $\delta$-NbN
(Table III) agrees very well with the experimental value~\cite{Ralls}.
The calculated $T_c$ is zero for pure $\varepsilon$-NbN, $\delta^\prime$-NbN and WC-NbN (Table III).
When large band smearing widths are  used to simulate the effect of random substitutial disorder
on the N sublattice due to the slight N-deficiency present in NbN samples,
the $T_c$ could go up to 1.0$\sim$3.0 K (Table III). This result agrees quite well with the
earlier report\cite{Oya} that $\varepsilon$-NbN did not exhibit superconductivity down to 1.77 K\cite{Oya}
but disagrees with the recent report\cite{YZou} that the $T_c$ of $\varepsilon$-NbN phase
in the samples with mixed $\varepsilon$-NbN and $\delta$-NbN phases is about 11.6 K.
To resolve this contraversy, we believe that further experiments on the single-phase samples
of $\varepsilon$-NbN would be helpful.

Finally, the calculated relativistic band structures reveal that all four NbN polytypes are
topological metals. In particular, both $\varepsilon$-NbN, $\delta^\prime$-NbN and $\delta$-NbN are, respectively, type-I and type-II
Dirac metals, which would exhibit novel quantum phenomena such as negative and anisotropic
magneto-transports~\cite{Liang14,Xiong15,Guo16} and topological phase transitions~\cite{Wang12}.
Furthermore, WC-NbN is an emergent topological metal that has exotic triply degenerate nodes.\cite{Bian,Lv17}
Therefore, all the four NbN polytypes should be hard superconductors with nontrivial band topology.
This suggests that the NbN polytypes would provide a valuable material platform
for studying fascinating phenomena arising from the interplay of band topology
and superconductivity.~\cite{Nandkishore12,Yang14b,Li18}

\section*{Acknowledgments}
The authors acknowledge support from the Ministry of Science and Technology and the Academia Sinica
of The R.O.C. as well as the NCTS and the Kenda Foundation in Taiwan.

\begin{figure}
\centering
\includegraphics[width=80mm]{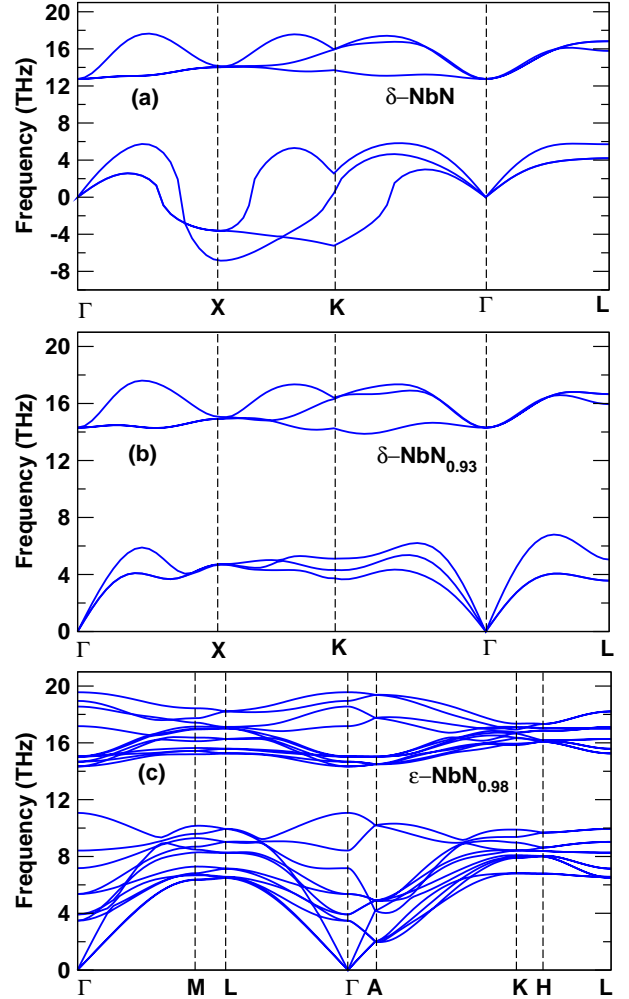}
\caption{Phonon dispersion relations of (a) pure $\delta$-NbN, (b) $\delta$-NbN$_{0.93}$ with N vacancies
and (c) $\varepsilon$-NbN$_{0.98}$ with N vacancies. The smearing width $\sigma = 0.02$ Ry is used in the calculations.
}
\end{figure}

\section*{APPENDIX: Possible effects of N deficiency on phonon dispersion}
Figure 7(a) shows the phonon dispersion of pure $\delta$-NbN with the standard electronic 
smearing width of $\sigma = 0.02$ Ry. The phonon dispersion shows imaginary frequency phonon modes at X and K points. 
This indicates that pure $\delta$-NbN is unstable. Experimentally, $\delta$-NbN phase could only 
be prepared with a small N deficiency\cite{Christ,XJChen}. 
One way to investigate the effect of the N deficiency is to calculate the phonon dispersion within the
virtual crystal approximation (VCA), i.e., the small N deficiency ($x$) in NbN$_{1-x}$ 
is simulated by a small reduction ($7x$) in the number of valence electrons. 
The phonon dispersion calculated based on the VCA for  $\delta$-NbN$_{0.93}$ and $\varepsilon$-NbN$_{0.98}$ 
are displayed in Figs. 7(b) and 7(c), respectively. Figure 7(b) shows that the soft phonon modes disappear
in $\delta$-NbN$_{0.93}$ while Fig. 7(c) indicates that the phonon dispersion of $\varepsilon$-NbN$_{0.98}$ 
is almost identical to that of pure $\varepsilon$-NbN [Fig. 5(b)]. 

In N-deficient $\delta$-NbN$_{1-x}$, N atoms and N vacancies (V$_N$) on the N sublattice are randomly distributed. 
The main effect of the disorder due to this random distribution of N and V$_N$
on electronic energy bands is a larger band smearing.\cite{Olifan} Therefore,
the effect of the disorder due to V$_N$ in NbN$_{1-x}$ could be taken into account by
using a large electronic smearing width in the calculations.\cite{Olifan}
The phonon dispersion relations calculated using a larger smearing value of $\sigma = 0.10$ Ry
for $\varepsilon$-NbN, WC-NbN and $\delta^\prime$-NbN are plotted in Figs. 8(a), 8(b) and 8(c), respectively.
Figures 5 and 8 show that the phonon dispersions calculated using the two different $\sigma$ values of
0.02 and 0.10 Ry are nearly the same. Therefore, we may conclude that although the small
N deficiency stabilizes the cubic  $\delta$-NbN by removing the imaginary 
frequency phonon modes at X and K points, it has negligible effects on the phonon dispersion 
in the hexagonal NbN polytypes ($\varepsilon$-NbN, WC-NbN and $\delta^\prime$-NbN).

\begin{figure}
\centering
\includegraphics[width=80mm]{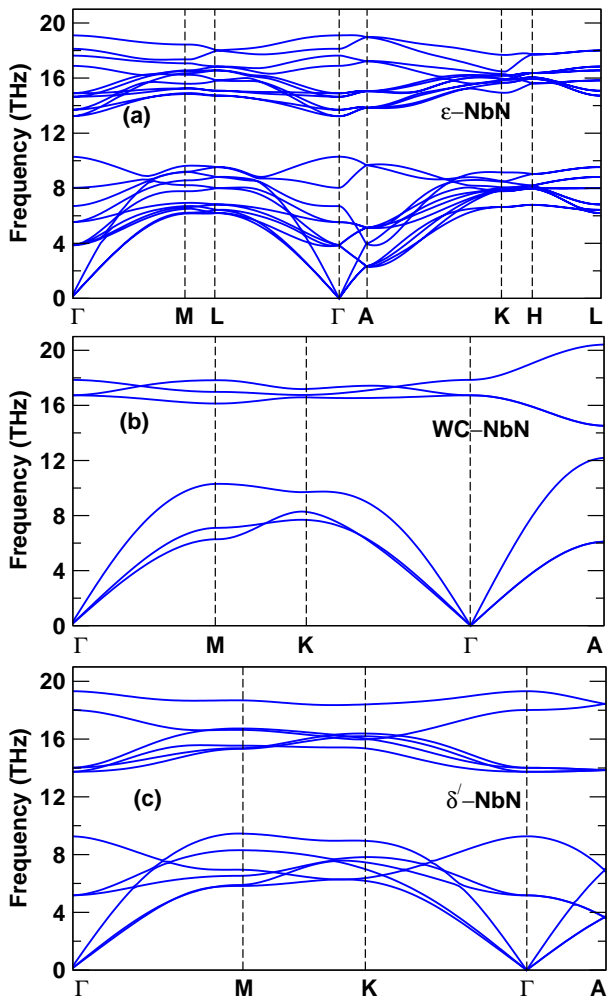}
\caption{Phonon dispersion relations of (a) $\varepsilon$-NbN, (b) WC-NbN and (c) $\delta^\prime$-NbN.
The plots are the same as that in Fig. 5 except that the smearing width $\sigma = 0.10$ Ry is used here.
}
\end{figure}


\end{document}